# Critical phenomena of nano phase evolution in a first order transition.

Yongseong Choi[1], David J. Keavney[1], Martin V. Holt[2], Vojtěch Uhlíř[3], Dario Arena[4], Eric E. Fullerton[3], Philip J. Ryan[1*] and Jong-Woo Kim[1&]

*[1] X-ray Science Division, Argonne National Laboratory, Argonne, IL, 60439, USA*

*[2] Center for Nano-Materials, Argonne National Laboratory, Argonne, IL, 60439, USA*

*[3] Center for Magnetic Recording Research, University of California, San Diego La Jolla, CA, 92093, USA*

*[4] National Synchrotron Light Source, Brookhaven National Laboratory, Upton, NY, 11973, USA*

In a first-order phase transition, a system transforms discretely from one state to another, however these transitions are often observed displaying continuous behavior. To understand this nature, it is essential to probe how the emergent phase nucleates, interacts and evolves with the initial phase across the transition at microscopic scales. Here, the proto-typical first-order magneto-structural transition in FeRh is used to investigate these phenomena. We find that the temperature evolution of the final phase exhibits critical behavior. Furthermore, a difference between the structure and magnetic transition temperatures reveals a *novel intermediate phase* created from the interface between the initial and nucleated final states. This emergent phase, characterized by its lack of spin order due to the competition between the antiferromagnetic and ferromagnetic interactions, leads to suppression of the dynamic aspect of the transition, generating a *static* mixed-phase-morphology. Understanding and controlling the transition process at this spatial scale is critical to optimizing functional device capabilities.

---

Correspondance to: * pryan@aps.anl.gov, & jwkim@aps.anl.gov

## Introduction

The physics of first-order phase transitions encompasses a broad range of technologically useful and fundamentally interesting phenomena across many scientific areas, such as liquid-solid transitions, magnetic ordering, superconductivity, reversal dynamics, and structural phase transitions. These transitions are typically driven by the imbalance between the free energies associated with the new phase and surface area of the interphase boundary, resulting in domain nucleation and growth behaviors.

A particularly interesting class is the first-order magneto-structural (MS) transition, which involves coupled degrees of freedom that lead to rich phenomenology of potential interest for device applications (1,2,3). The CsCl-structure ordered alloy $Fe_{50}Rh_{50}$ demonstrates a MS transition from an AFM to a FM state upon heating from room temperature to above ~370 K (1,2,15-18) with a temperature hysteresis of about 10 K between heating and cooling cycles, accompanied by a volume increase of 1-2%, a reduction in resistivity and a large change in entropy. This transition can be driven by several external parameters such as temperature, magnetic field or pressure. In particular, the transition temperature displays sensitivity to external magnetic fields of -9 K/T, indicating the truly coupled nature of the structural and magnetic degrees of freedom (4,5). Therefore, the $Fe_{50}Rh_{50}$ alloy is a test-bed for exploring the interplay of structural, magnetic and electronic phase transitions, and offers an exceptional opportunity to understand these processes. Complementary qualities including large magneto-resistance and magneto caloric effects are critical features in the development of new technologies for magnetic sensors, memories and refrigeration (1,6-8).

Compared with the typical sharp character, broad first order transitions are often observed as a result of sample inhomogeneity, which, by way of defects act as catalysts causing surface or interface melting resulting in continuous transitions (9-11). Local disorders often generate nucleation sites initiating the emerging phase but may also impede subsequent domain growth. These defects leave residual metastable phases post-transition resulting in glassy behavior (12-14). As such, understanding the disorder-driven heterogeneous nucleation process is crucial to control potential functionality based upon the transition itself. Moreover, the continual scaling of device architecture will require better understanding of the implications defects have upon microscopic physical properties.

In this work, we study the MS properties of epitaxial FeRh thin films grown on MgO (001) substrates (shown schematically in Fig. 1) which exhibit a broad first order MS transition at ~370 K (Supplemental figures S1 & S3). Employing nanometer resolved x-ray contrast imaging we monitored the critical growth behavior occurring in the first order structural transition of the epitaxial film. In conjunction, ferromagnetic x-ray microscopy illustrates the domain growth of the FM state while the additional combination of *in situ* magnetic dichroic spectroscopy and x-ray diffraction (XRD) using common thermometry reveals the element-specific development of magnetic and structural order. Two atypical phenomena are presented, first, the temporally *static* coexisting phase morphology during the MS transition and second and most remarkable, *critical behavior* of the phase growth process. Here we present evidence describing an intermediate phase emerging from the interface between the coexisting and competing magnetic AFM and FM spin states. This emergent state generates an environment akin to a 'wetting layer' surrounding the emerging nucleated phase, seeded by intrinsic defects. The intermediate state extends from the interphase boundary and is comprised of a spin frustrated magnetic regime through the competitive magnetic interactions. Such an anomalous spin state suppresses dynamic growth inducing critical behavior.

**Direct observation of nano-scale structural phase evolution**

Previously, x-ray diffraction studies observed phase coexistence of both high and low temperature states during the MS transition (19). The discrete change of the lattice parameter and the large temperature hysteresis are consistent with first-order behavior (Fig. S1). Nano-scale x-ray diffraction microscopy enables spatial mapping of both states utilizing distinct Bragg reflections (20-22). The nucleation and growth process was then directly mapped as a function of temperature (Fig 2).

The emerging phase begins with a droplet shape indicating nucleation is formed by point defects. Counterintuitive for a first order transition, which generally displays dynamic growth after nucleation, temporally stable but thermally driven domain size evolution follows a critical behavior, $1/(T_t-T)$ where $T_t$ is the temperature of the transition. This phenomenon is reminiscent of critical behavior of a correlation length for a second-order transition with a length scale of several hundred nanometers. Large-scale correlation lengths have been observed in heterogeneous nucleation processes with both second and weak first-order transitions. In these cases, elastic deformation from defect centers can generate strain fields that extend over large distances, creating large-scale correlations prior to second-order transitions (23-25). However, the observed long-range critical behavior is extraordinary in this case of a strong first order transition.

**Intermediate phase**

In addition to strain, a magnetic exchange field also contributes to the free energy landscape of the FM-AFM interfacial region. In order to understand the nature of the MS transition, characterization of the magnetic response corresponding to the structural transition is required. The FM signature is measured with x-ray magnetic circular dichroism (XMCD) at both the Rh $L_2$ and Fe K edges while the lattice expansion and contraction is measured concurrently by XRD, assuring accurate comparative thermometry (Supplemental Fig. S3). In this way, the element specific FM response is directly compared with the structural evolution throughout the transition. A clear thermal gap is observed between the magnetic and structural behaviors in Fig. 3a. Both the Fe and Rh FM transition temperatures are higher than the corresponding structural transition. Previous neutron measurements demonstrated that the loss of AFM order coincides with the lattice expansion in the heating process (2). Considering this, our result reveals a gap of ~6 – 7 K in the magnetic ordering, in which an additional phase emerges that is neither FM nor AFM concomitant with the transition and thus phase coexistence. It is this intermediate state that introduces second-order behavior to the MS transition by generating a passivation barrier between the coexisting and competing spin phases.

At 380 K (Fig. 3a), a small fraction of the film displays FM order, while roughly half the film has undergone the structural transition (26). This volume difference between the structural and magnetic components reveals the intermediate phase, which exhibits lattice expansion but not FM ordering, as illustrated in Fig. 3b. We consider the structural change is driven by the onset of FM interactions, however in this phase, the subsequent lattice expansion itself does not complete the MS transition. Additionally striking is the FM ordering of the Rh moments before the Fe, providing evidence that the FM interaction is mediated by the Rh spin. Previous photoemission spectroscopy results revealed that the electronic structure change across the MS transition is dominated by the Rh 4d character (27). This is consistent with the significant role Rh is reported to play (16,28,29). Changes to the Rh electronic configuration with increasing temperature enhance the local moments subsequently driving the MS transition. The intermediate phase encapsulates the final FM nucleation sites, illustrated in Fig. 3b. This generates 'perfect wetting', a condition that forms to reduce the interfacial energy by replacing the boundary between the first and second phases even if the wetting state is not an independently stable phase (30). In solid-liquid transitions surface melting is normally first order in nature but becomes second order due to the wetting process. Here we observe an analogous phenomenon at the boundary of an emerging phase, driven by defect nucleation, which is confirmed by the return point memory effect of thermal cycling in both structural and magnetic phase domain imaging. (Supplemental Figure S4)

## Microscopic magnetic phase transition

In order to observe the Fe FM domain evolution across the transition, we employed XMCD photoemission electron microscopy (PEEM). Using the strong dichroic effect at the Fe $L_3$ edge with circularly polarized radiation, a magnetic domain image is obtained from the difference of left and right circular images. Thus, the contrast is related to the spin direction with respect to the *k*-vector of the incoming circularly polarized beam, Fig S6. The magnetic imaging, shown in Fig. 4, indicates that through the magnetic transition, morphological dichroic contrast is clearly observed within the region of the single ferromagnetic domain (at 393 K) specified by the boxed area. At the midpoint of the XMCD-PEEM transition (~375K), the FeRh has almost completed the structural transition (Fig. 3a) and AFM order is no longer preserved (2). At this juncture the persistent blue regions represent the absence of dichroic contrast and is therefore non-FM, indicating a significant volume of spin disorder corresponding to the intermediate phase. To follow these domains more closely, several 200 × 200 $nm^2$ regions of interest are indicated, and their pixel averages plotted in Fig. 4c. The regions of interest each present a continuous transitional behavior with a similar degree of contrast, but with varying transition temperatures. This shows that the FM phase grows gradually, even within a 200 × 200 $nm^2$ field of view (31). The variant temperatures for the selected volumes illustrate how FM onset is spatially inhomogeneous. Consequently the FM phase domain initiates from dispersed nucleation points and grows with temperature. We note that the FM order displays a degree of return point memory (Fig. S4), indicating that the dispersion of the nucleation points is not random but seeded by intrinsic defects generating magnetic pinning.

## Discussion

Systems with a first-order transition show anomalous behavior at defect and impurity positions. Surfaces are ubiquitous defects modifying the free energy landscape and a wetting layer forming at the surface often undergoes a transition with several critical phenomena characterized by critical exponents. This characteristic behavior is observed at the interface between coexisting magnetic phases during the magneto-structural transition in FeRh. In this case, the wetting phase is generated by intrinsic defect driven nucleation events, as opposed to extrinsic conditions such as the surface (as presented by magnetic reflectivity in Fig. S7). Emerging FM domains are completely surrounded by the intermediate wetting phase. This interfacial state leads to a dynamical growth process evolving into a temporally stable state by creating a passivation barrier between two typically incompatible states. The intermediate phase presents universal criticality, as seen in surface effects upon first order transitions (32,33). Furthermore, magnetic 'glassy' behavior arising from AFM-FM competition is likely due to an intermediate wetting phase (3,7,8). The existence of the intermediate phase is an essential part of the behavior of magneto-structural transitions and an important factor to control their functional properties. Engineering deliberate local defects by doping, irradiation or microscopic patterning will allow greater control of this functional behavior.

## Methods

Epitaxial 50-nm-thick FeRh films were grown on MgO (100) substrates at 450 °C and an argon pressure of 1.5 mTorr by dc magnetron sputtering using an equiatomic target. The films were post-annealed at 850 °C for 2 hours.. The crystallographic orientation is such that [100] direction of FeRh aligns with [110] direction of MgO (i.e. the FeRh crystal lattice is rotated by 45 degrees – Fig. 1).

Nanoscale X-ray Diffraction Microscopy experiments were performed using the Hard X-ray Nanoprobe (HXN) of the Center for Nanoscale Materials (CNM) at sector 26-ID-C of the Advanced Photon Source, Argonne National Laboratory. The monochromatic incident X-ray beam (photon energy 10.0 keV, $\lambda$=1.2398 Å) was focused by a Fresnel zone plate yielding a ~30 nm beam size at the sample.

Soft x-ray energy magnetic circular dichroic (XMCD) images were taken at the Advanced Photon Source beamline 4-ID-C using an Omicron photoemission electron microscope (PEEM). This instrument uses electrostatic electron optics to image the emitted secondary electrons from the FeRh surface with a typical spatial resolution of ~100 nm. The beamline uses a helical undulator and a spherical grating monochromator tuned to the Fe $L_3$ resonance with a typical bandwidth of 0.05%. Difference XMCD images were obtained by acquiring separate exposures with LCP and RCP polarized radiation. The sample temperature was controlled using a filament mounted behind the sample and monitored with a thermocouple mounted on the sample holder.

Intermediate and hard x-ray XMCD measurements were performed at the 4-ID-D beamline. A circularly polarized x-ray beam was generated with a diamond phase retarder. A magnetic field of ±500 Oe was applied parallel to the sample plane. XMCD spectra were recorded at the Fe K edge in fluorescence mode using an energy dispersive detector. Magnetic dichroic signals at the Rh $L_2$ edge were recorded in reflectivity mode using asymmetry ratio.

**Acknowledgements:** We would like to thank Stephanie Moyerman for help on sample fabrication. Work at Argonne is supported by the US Department of Energy (DOE), Office of Science, Office of Basic Energy Sciences, under Contract No. DE-AC-02-06CH11357. Work at UCSD was supported by DOE-BES Grant Number: DE-SC0003678.

**Figure Captions:**

Fig. 1: First order magneto-structural (MS) transition in a FeRh film grown on MgO(001). Cartoon illustration of the uniaxial expansion and spin reordering of strained FeRh films on MgO (001) through the transition from AFM to FM.

Fig. 2: Structural evolution of the first order MS transition using x-ray nano diffraction (N-XRD) contrast imaging. a) The experimental schematic of N-XRD setup. b) Temperature dependence of domain sizes for both AFM and FM phases. The red dot and blue square represent FM and AFM domain sizes extracted from the nucleation points indicated by the arrows in Fig. 2 (c) respectively. The domain growth behavior follows $1/(T - T_t)$, with T = sample temperature and $T_t$ = transition temperature, 373.5 K. c) A temperature series of diffraction images through the MS transition. The scanned area is $1.5 \times 1.5$ μm$^2$ and consists of $30 \times 30$ pixels with a 30 nm cross-section incident beam.

Fig. 3: Temperature evolution of the structural and magnetic parameters through the MS transition. a) Comparative plot of the structural and magnetic thermal evolution (both Fe and Rh) from AFM to FM upon heating employing XRD and XMCD. b) A schematic illustration of the four states of the transition, from left to right, AFM Fe spin ordering with contracted volume, spin disorder with expanded volume, Rh only FM order with expanded volume and finally completed transition with both Fe and Rh FM collinear ordering with expanded lattice. c) Temperature comparison in asymmetry ratio of the magnetic reflectivity near the Rh $L_2$ edge. d) Temperature comparison the of Fe K edge XMCD signal below and above the transition.

Figure 4: Temperature dependence of Fe ferromagnetic domain images using XMCD-PEEM. a) A series of XMCD-PEEM images acquired in zero field at the Fe $L_3$ edge with increasing temperature. b) Schematic of the experimental measurement illustrating the electron optics. c) Magnified view of the selected (boxed) region in a). d) A plot of magnetic dichroic values derived from the standard deviation of the XMCD-PEEM images. This is proportional to the average FM magnetization. The XMCD signal for five 200 x 200 $nm^2$ ROIs as shown by the numbered boxes in a). For each of these small regions, the phase transition is ~7 K wide, with variations in the transition temperature of several degrees. While the XMCD signal is half of the maximum, almost the entire system is in the structural high temperature phase. At the corresponding temperature (375 K), the PEEM image shows a similar disparity of non-FM areas. Noting that in this experimental geometry all FM domains elicit a dichroic signal (Fig. S6) and that complete suppression of AFM order occurs with lattice expansion (2), therefore the 'missing' dichroic signal in the PEEM image at this point, indicates that a significant volume are neither AFM nor FM but represent the emerging spin disorder (SD) phase.

Fig 1.

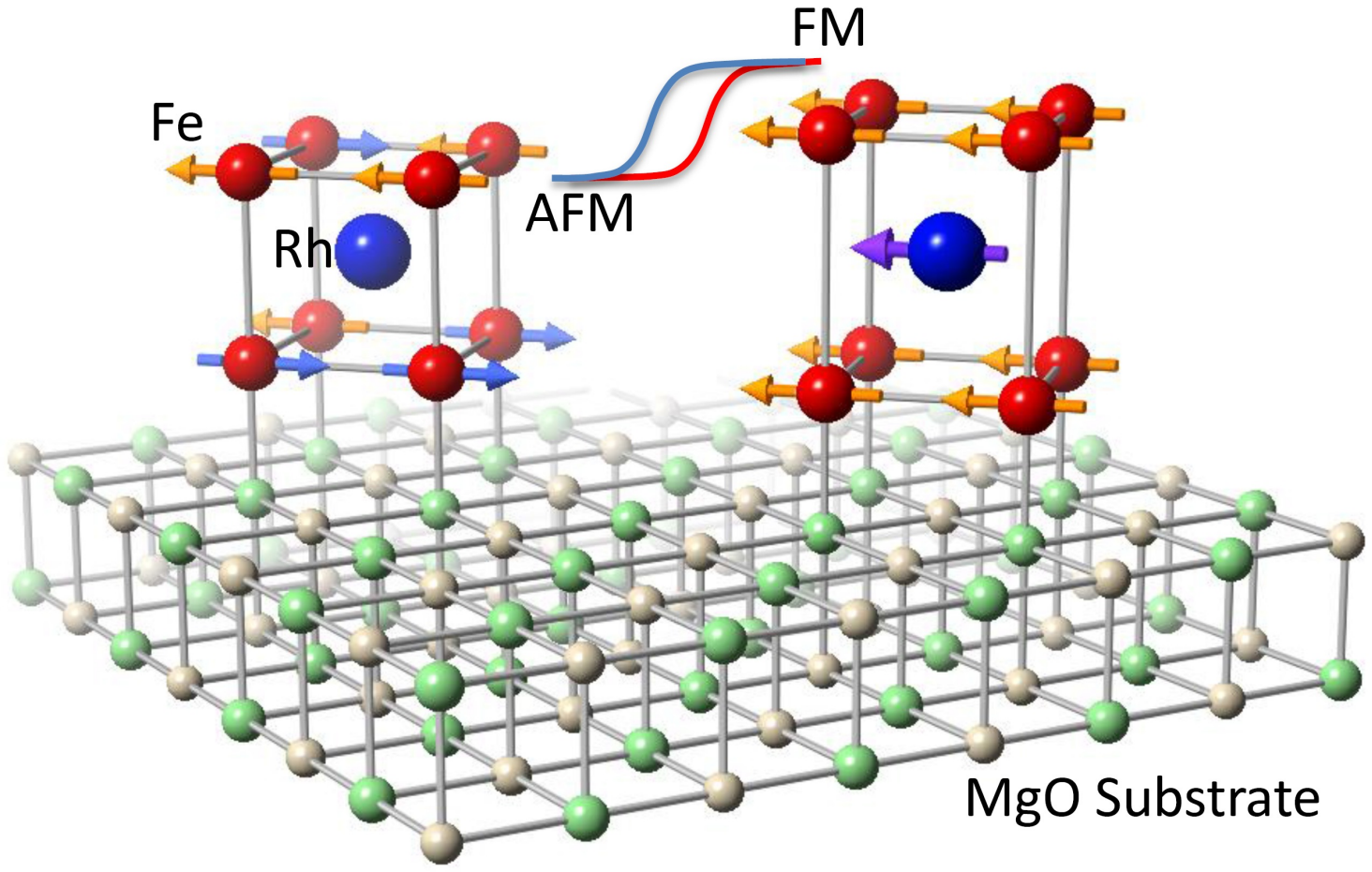

Fig. 2

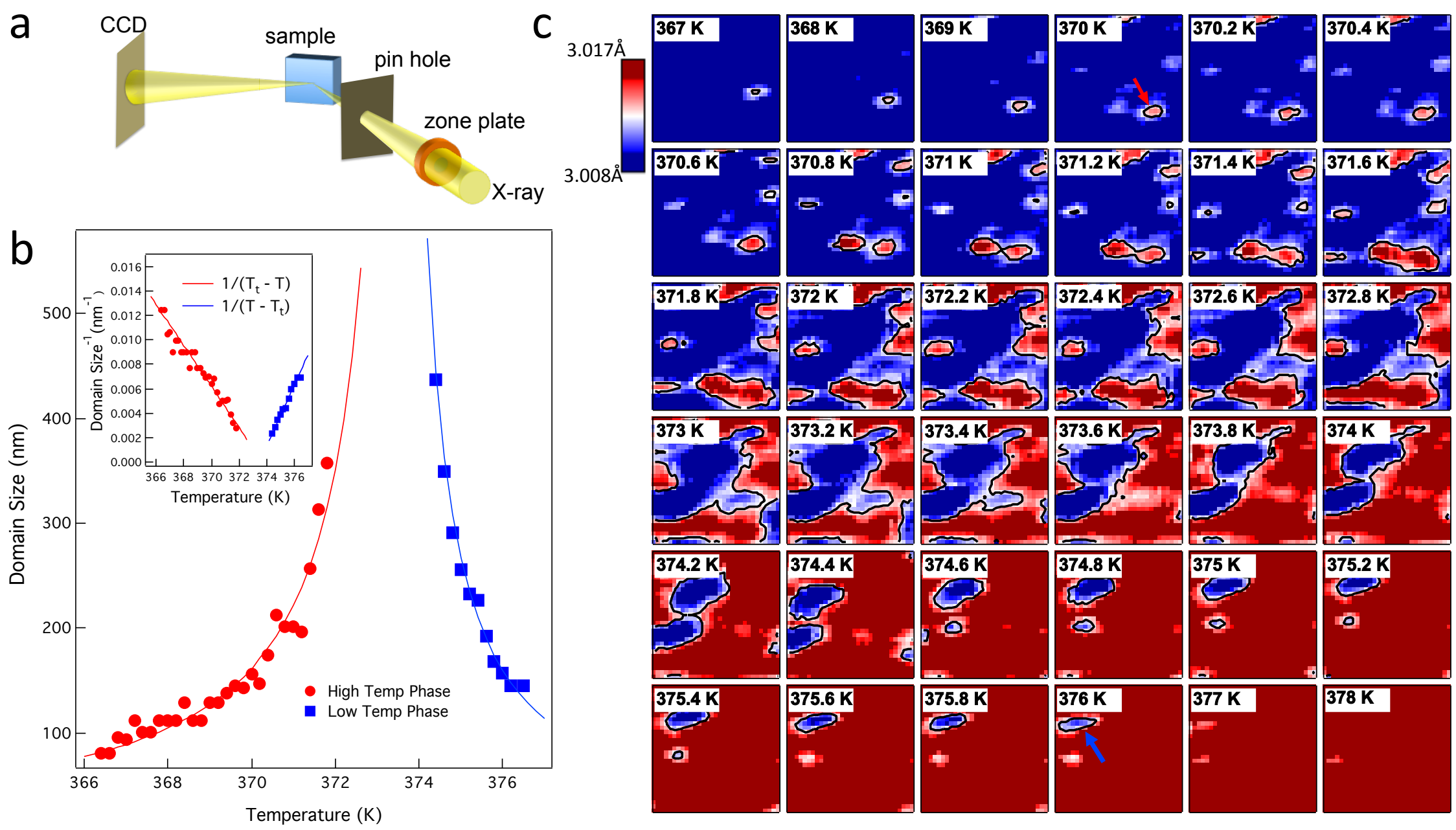

Fig. 3.

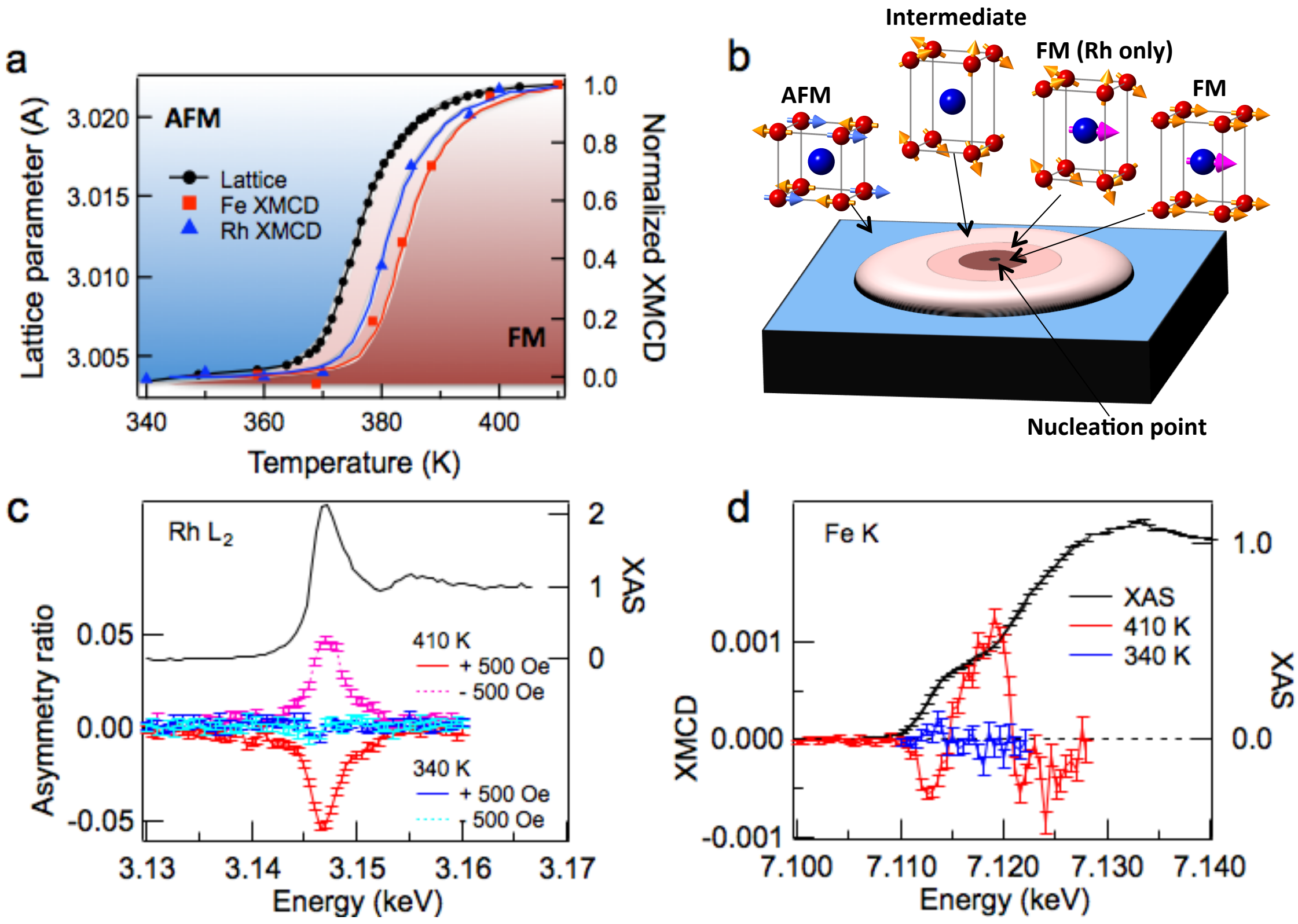

**Fig. 4**

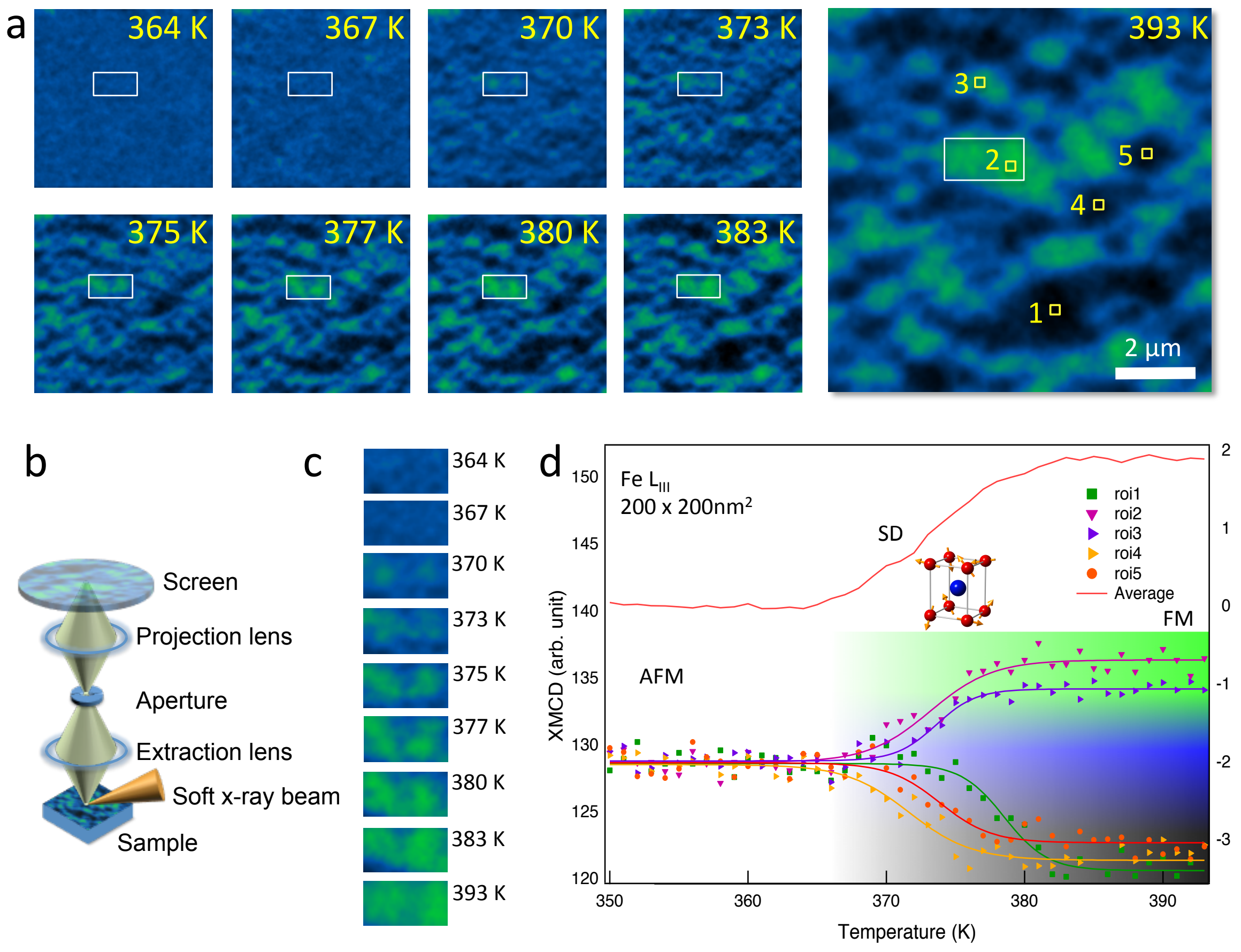

**Supplemental Information:**

# Critical phenomena of nano phase evolution in a first order transition.

Yongseong Choi[1], David J. Keavney[1], Martin V. Holt[2], Vojtěch Uhlíř[3], Dario Arena[4], Eric E. Fullerton[3], Philip J. Ryan[1*] and Jong-Woo Kim[1&]

*[1] X-ray Science Division, Argonne National Laboratory, Argonne, IL, 60439, USA*

*[2] Center for Nano-Materials, Argonne National Laboratory, Argonne, IL, 60439, USA*

*[3] Center of Magnetic Recording Research, University of California, San Diego La Jolla, CA, 92093, USA*

*[4] National Synchrotron Light Source, Brookhaven National Laboratory, Upton, NY, 11973, USA*

## **Structural transition measured by XRD**

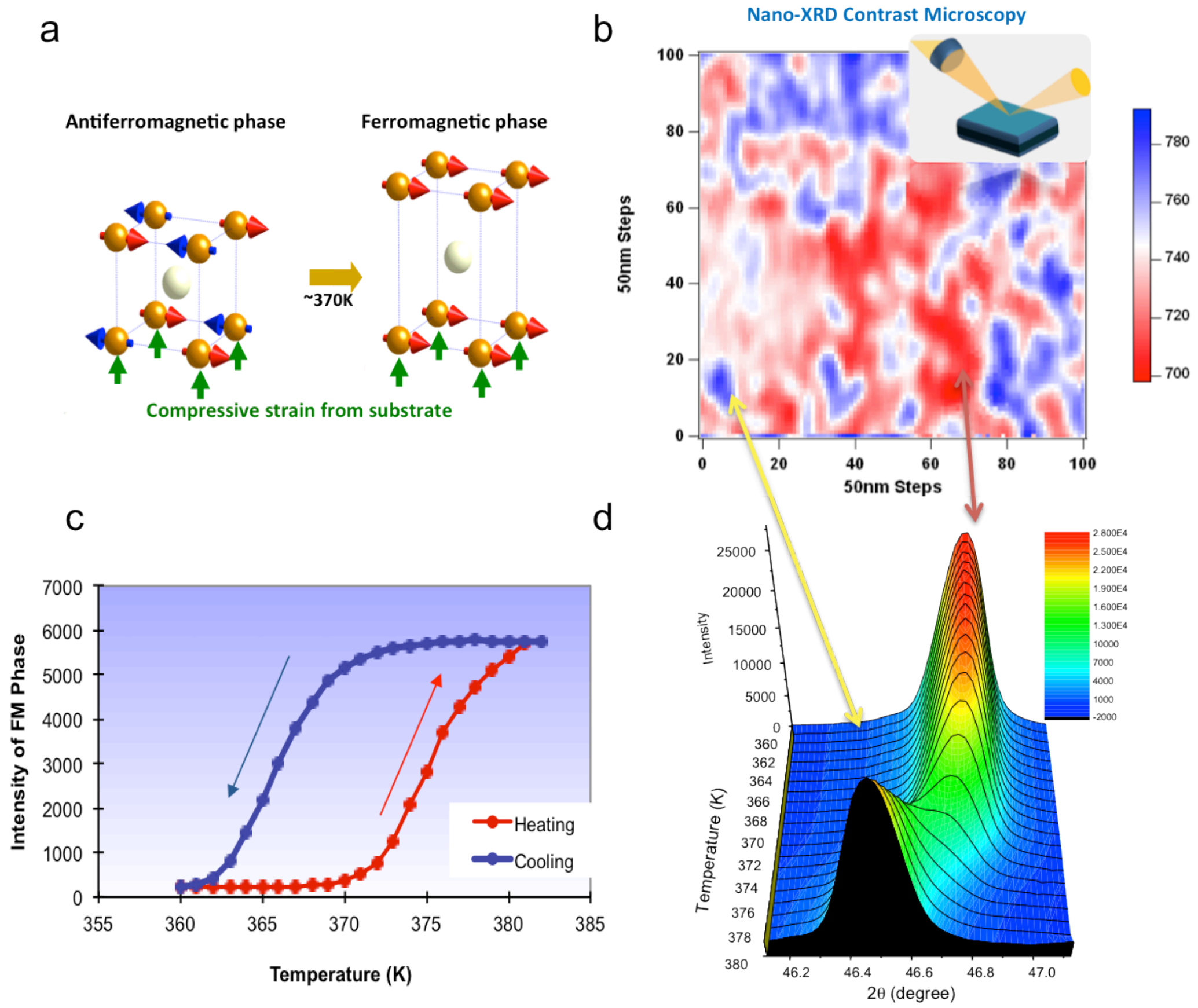

**Fig. S1.**

High resolution XRD was performed at 6-ID-B beamline at the Advanced Photon Source, (APS), while N-XRD microscopy was carried out at beamline 26-ID also at the APS. a) Illustrates the structural and magnetic order changes through the MS transition. b & d) Contrast diffraction imaging with respect to the distinct reflections in parallel diffraction. c) Change of the diffraction peak intensity for the higher temperature phase with both increasing and decreasing temperature showing the broad hysteretic behavior of the transition.

**Sample environment for concurrent XRD and XMCD measurements**

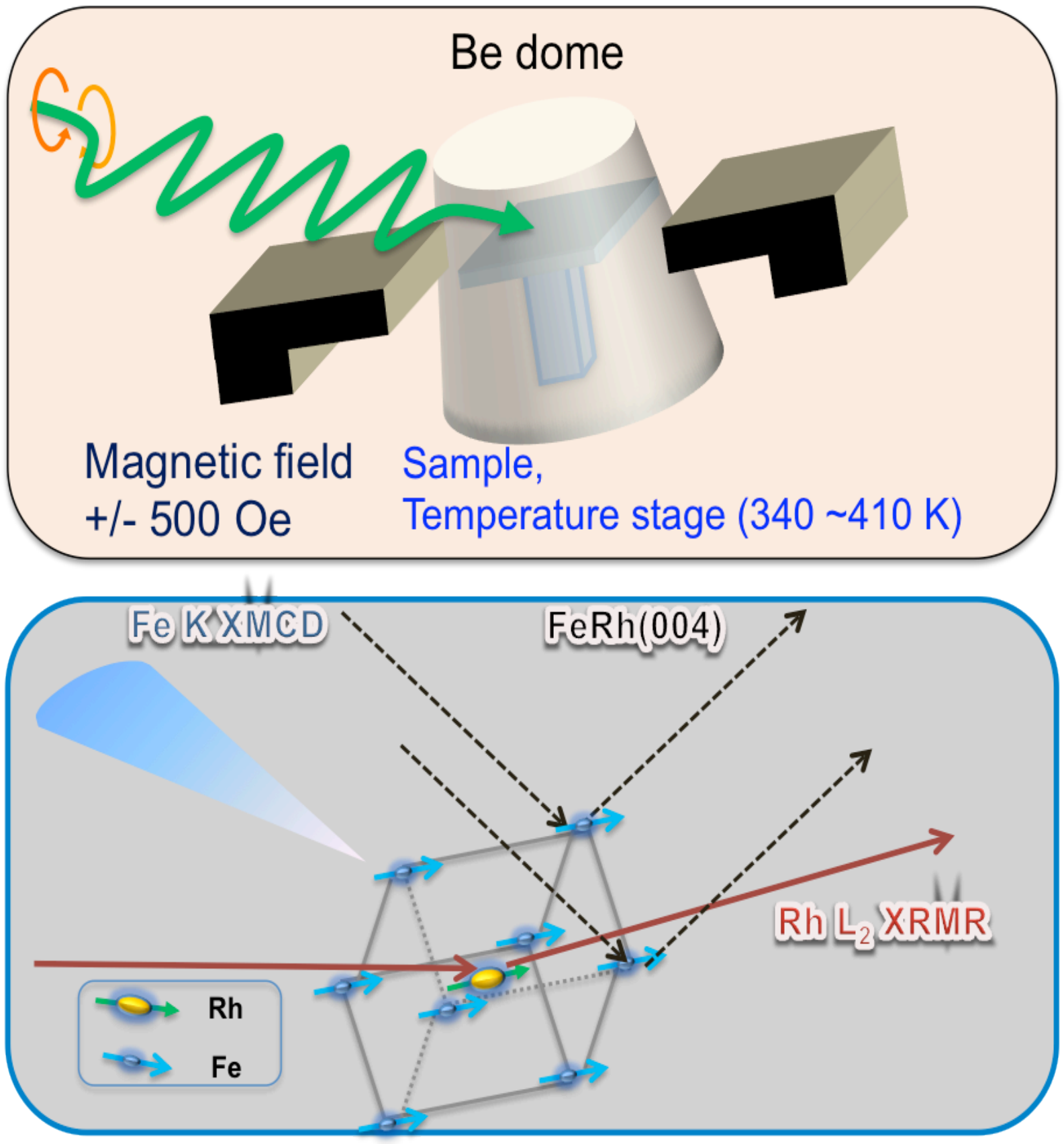

**Fig. S2.**

a) Cartoon model of *in-situ* sample environment allowing both the structure and magnetic parameters to be measured simultaneously.  b) Illustration of how the energy tunability of synchrotron radiation was employed to extract the structure and elementally resolved spin ordering with one sample environment. While the magnetic XMCD used Fe K and Rh $L_2$ edges, the lattice parameter was measured using 9.4 keV achieved using a higher harmonic of the source (the magnetic undulator insertion device (ID)).

**Hysteretic Effect in Both Fe and Rh magnetic behaviors compared with XRD**

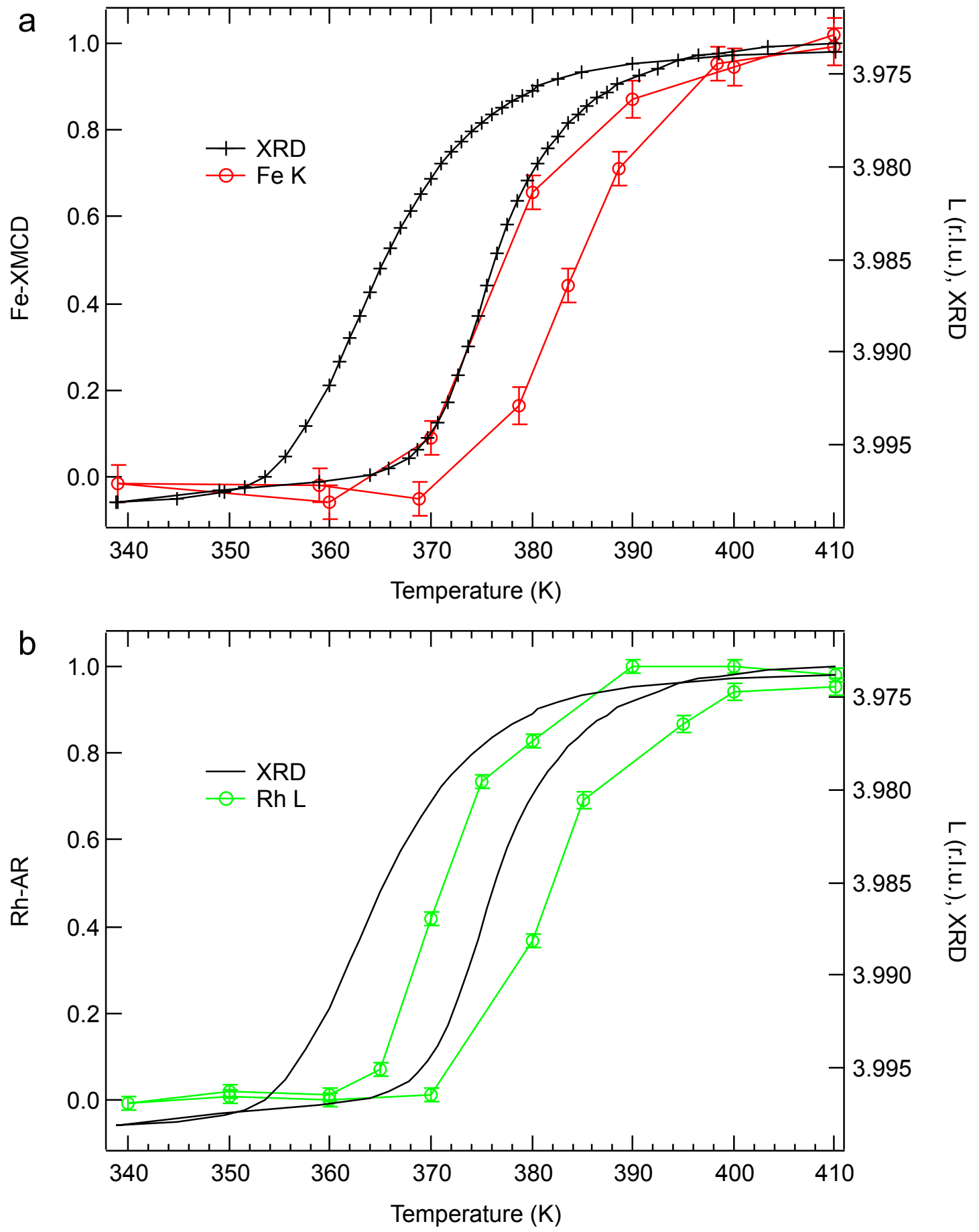

**Fig. S3.**

XMCD measurements were performed on FeRh film at the 4-ID-D beamline of the APS, Argonne National Laboratory. XMCD spectra were recorded at the Fe K edge in fluorescence mode using an energy dispersive detector. Magnetic dichroic signals at the Rh $L_2$ edge were recorded in reflectivity mode using asymmetric ratio.

## Return point memory effect

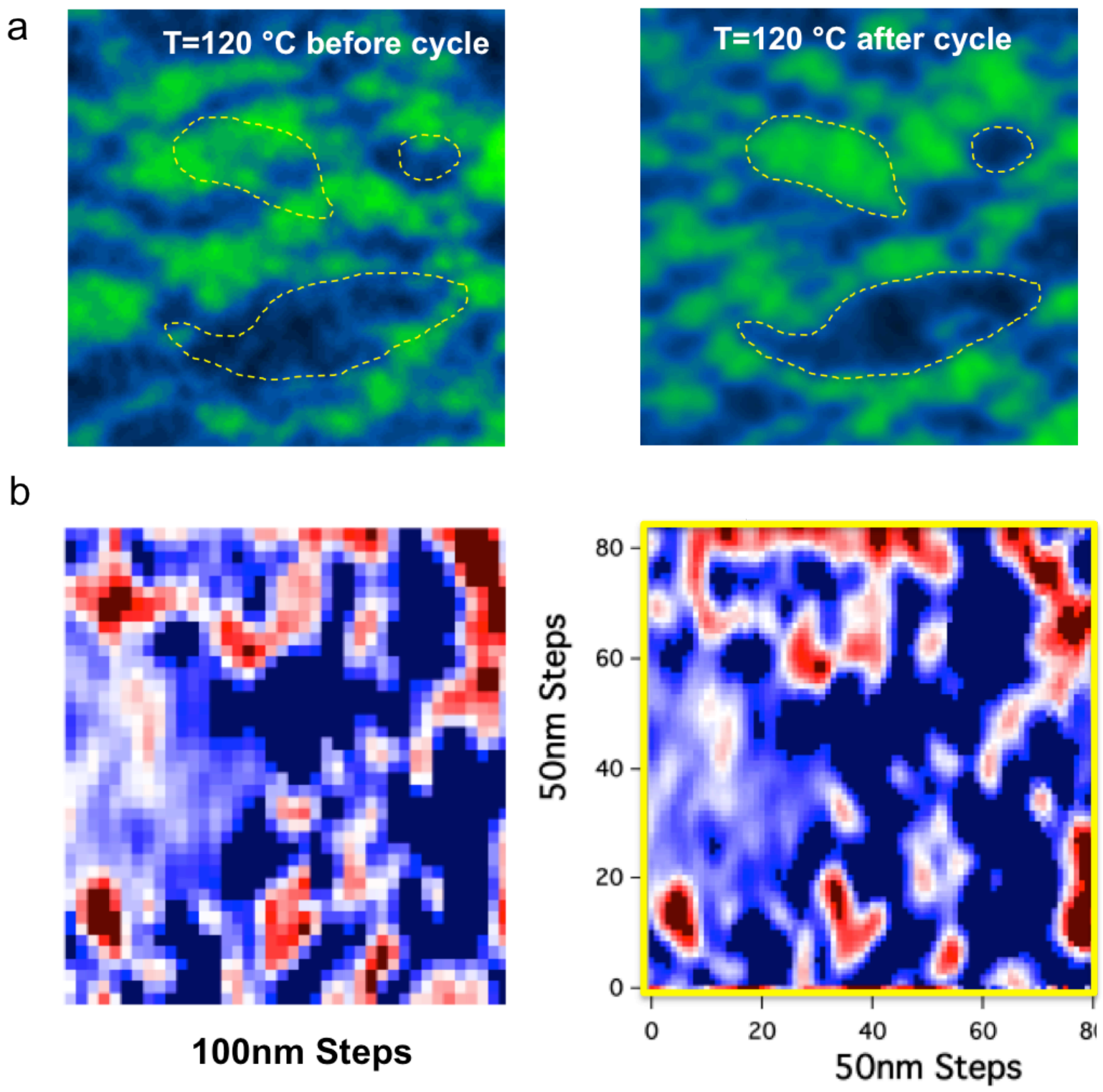

**Fig. S4.**

Both a) XMCD-PEEM and b) N-XRD indicate a significant degree of return point memory (RPM). Both data sets present the same position at the same temperature before and after a single cycle (PEEM :T = 393K, 9.6 × 9.6 μm$^2$, NXRD : T = 349 K, 4 × 4 μm$^2$). General shapes of the FM domains are preserved after cooling/heating cycles, although the smaller features tend not to be preserved. This suggests that the initial pinning sites initiating nucleation are generally the same with the growth affecting each cycle stochastically.

**XMCD PEEM – Full Temperature Cycle (Video)**

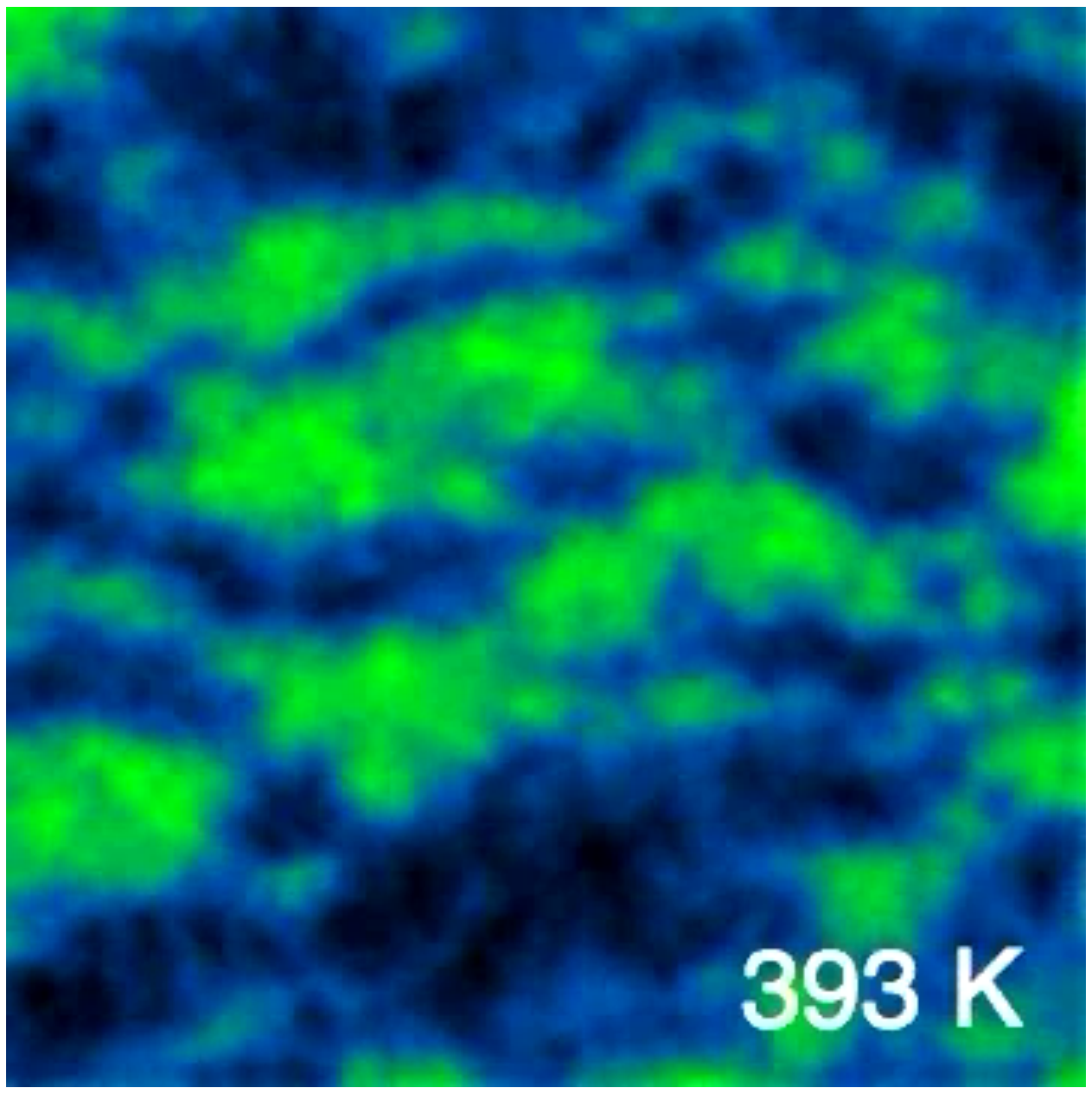

**Fig. S5.**

Beginning at the high temperature ferromagnetic phase 393K the temperature cycles below the transition to 330K and returns to the starting temperature.

**PEEM Image Domain Orientation.**

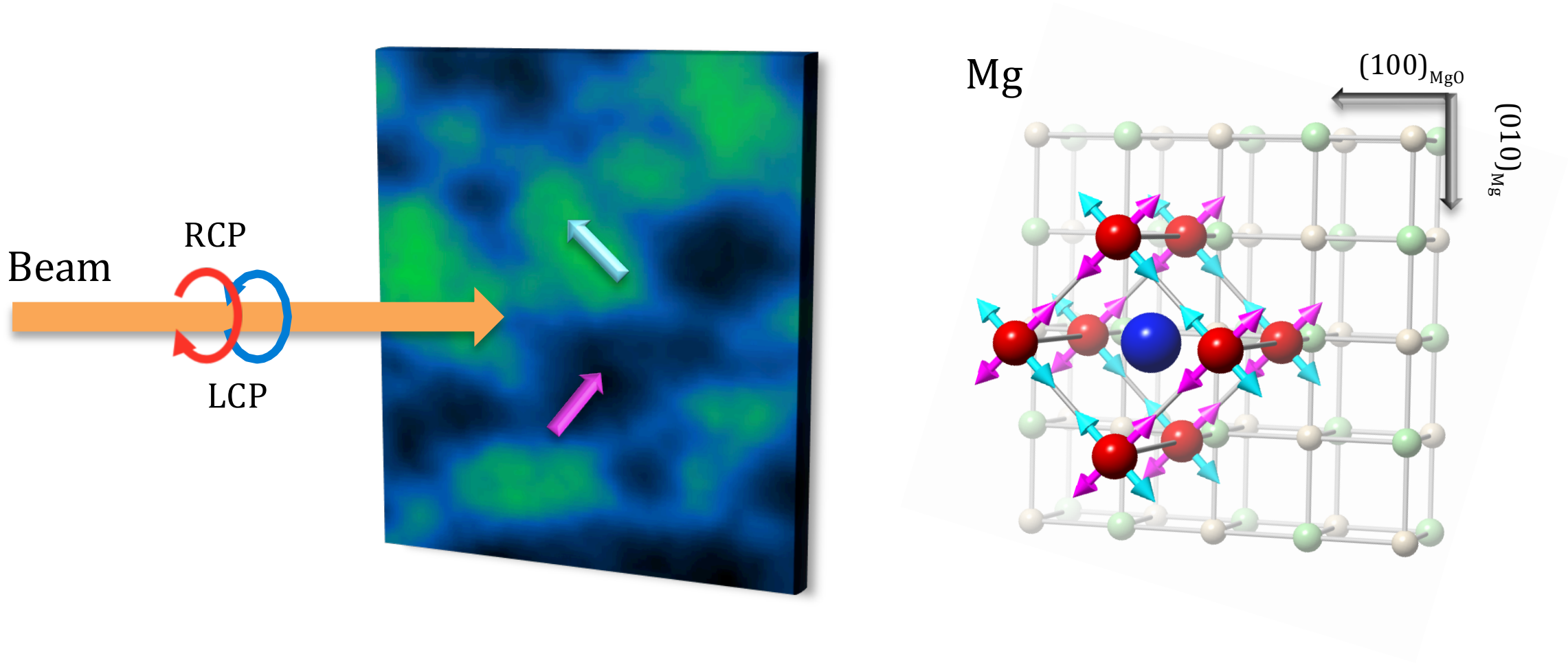

**Fig. S6.**

Illustration of FM spin orientation domain contrasts of the XMCD-PEEM. The FeRh is 45° rotated from the MgO substrate. The easy axis of the FM order in the FeRh is along the principal crystallographic axis <100>. The circularly polarized incident beam is 25° out-of-plane. Due to the 45° azimuthal orientation of the spins with respect to the incoming vector of the circularly polarized beam, the complete volume of the illuminated sample presents a dichroic signal as there will not be spin domains orthogonal to the plane generated by the incoming beam and sample normal.

**Temperature dependence of Rh magnetic depth profile**

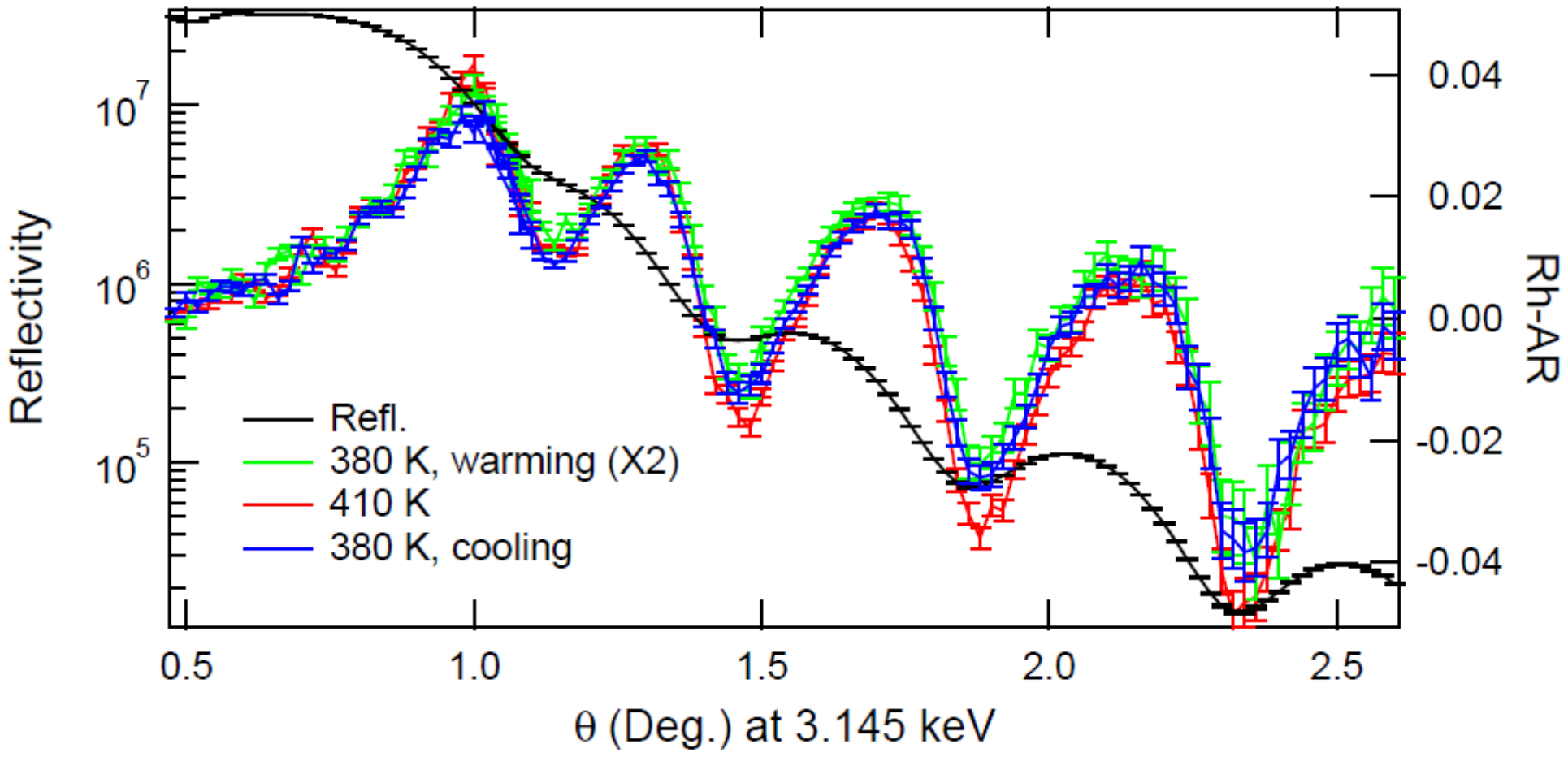

**Fig. S7.**

Asymmetry ratio (AR) in magnetic reflectivity curves near the Rh $L_2$ edge were measured during warming and cooling cycles as well as the maximum measurement temperature. The line-shape in AR is sensitive to the magnetic depth profile. At these temperatures, asymmetry ratio simply changes by a scaling factor without any significant change in line shape. This indicates that the phase domain growth occurs primarily in the plane of the film while in the surface normal direction the transition happens instantaneously. The similar magnetic depth profiles support that the observed temperature evolution is not due to surface/interfacial effects.

## H-Field dependence of magnetic transition temperature

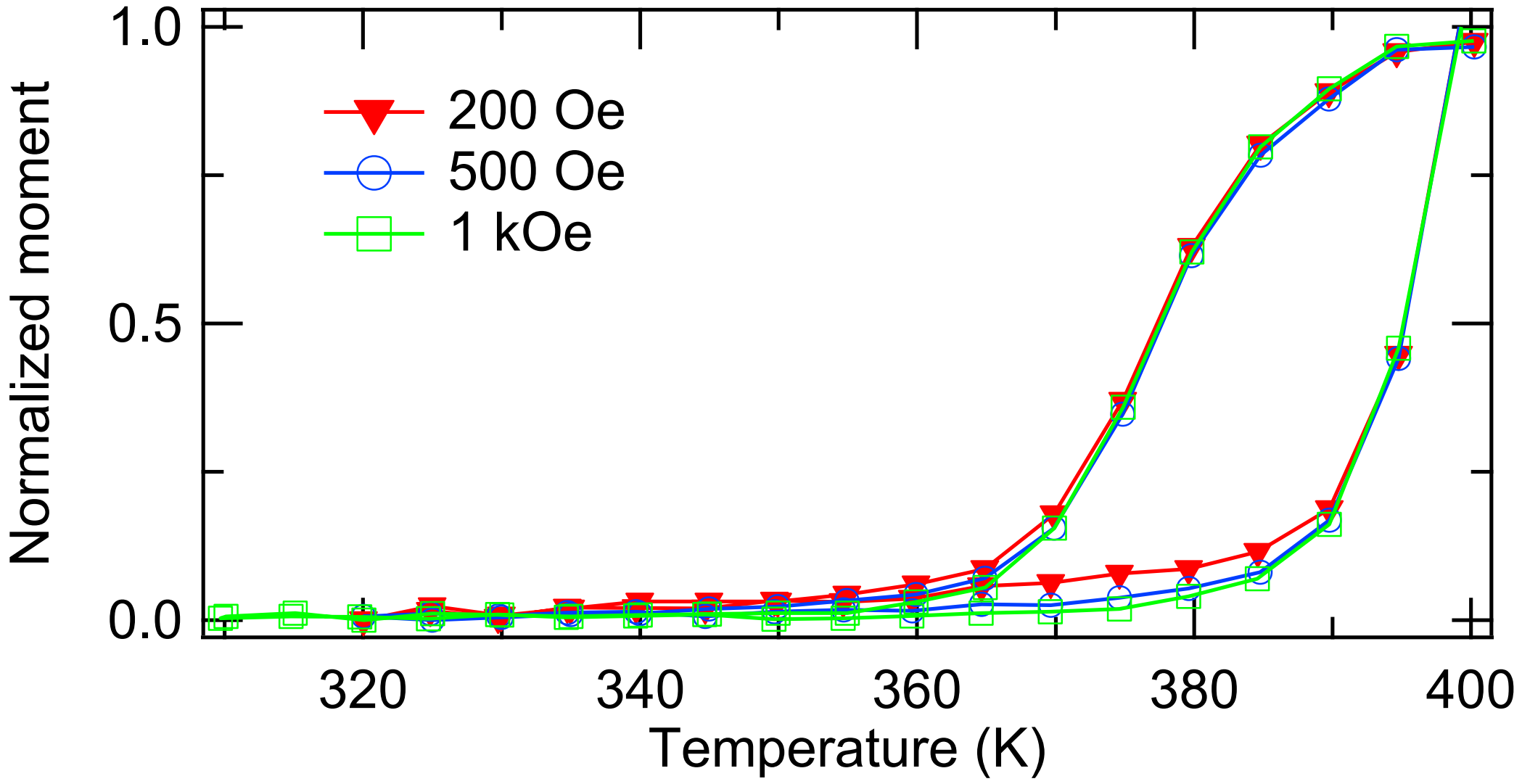

**Fig. S8.**

Vibrating sample magnetometry (VSM) temperature loops under select applied magnetic fields are shown. Each loop is normalized to the maximum value at 400 K. The hysteresis loops under 200, 500, and 1000 Oe have similar shapes.